# A Foundation for Stochastic Bandwidth Estimation of Networks with Random Service

Ralf Lübben   Markus Fidler
Institute of Communications Technology
Leibniz Universität Hannover

Jörg Liebeherr
Department of Electrical and Computer Engineering
University of Toronto

**Abstract**

We develop a stochastic foundation for bandwidth estimation of networks with random service, where bandwidth availability is expressed in terms of bounding functions with a defined violation probability. Exploiting properties of a stochastic max-plus algebra and system theory, the task of bandwidth estimation is formulated as inferring an unknown bounding function from measurements of probing traffic. We derive an estimation methodology that is based on iterative constant rate probes. Our solution provides evidence for the utility of packet trains for bandwidth estimation in the presence of variable cross traffic. Taking advantage of statistical methods, we show how our estimation method can be realized in practice, with adaptive train lengths of probe packets, probing rates, and replicated measurements required to achieve both high accuracy and confidence levels. We evaluate our method in a controlled testbed network, where we show the impact of cross traffic variability on the time-scales of service availability, and provide a comparison with existing bandwidth estimation tools.

## 1 Introduction

The objective of available bandwidth estimation is to infer the service offered by a network path from traffic measurements taken at end systems only. In recent years, available bandwidth estimation has attracted significant interest and a variety of measurement tools and techniques have been developed, e.g., [20–22, 32, 36, 37]. There has recently been a growing interest in obtaining a foundational understanding of the underlying problem, in particular considering the variability of cross traffic and the effects of multiple bottleneck links.

In bandwidth estimation methods, end systems exchange timestamped probe packets, and study the dispersion of these packets after they have traversed a network of nodes. The vast majority of estimation methods interpret the packet dispersion under the assumption that probe traffic flows through one or more fixed-capacity FIFO links that experience cross traffic [20, 32, 37]. The same FIFO assumption is found in most analytical studies of bandwidth estimation schemes [12, 18, 29–31, 33]. While FIFO queueing may be highly prevalent in wired network infrastructures today, limiting bandwidth estimation to such networks does not fully exploit the potential of network probing methodologies. This motivates the creation of a new foundation for bandwidth estimation for network environments, where FIFO assumptions are difficult to justify, such as wireless or multi-access networks [7].

A recent system-theoretic approach [28] presented an interpretation of bandwidth estimation that dispensed with the FIFO assumption for network elements. Instead, the network is viewed as a general time-invariant system where throughput and delays of traffic are governed by an unknown bounding function, referred to as *service curve*. (In the language of system theory, the service curve is the impulse response of a min-plus linear system [26].) While the long term rate of the service curve corresponds to the available bandwidth, the service curve encodes additional information, e.g., it can express bandwidth availability at different time scales and it can account for delays in the network. A particular strength of service curves is that they offer a natural extension to networks with several bottleneck links, by exploiting properties of the network calculus [10, 26]. Existing packet train probing schemes, similar to Pathload [21] and Pathchirp [36], were shown to be compatible with the system-theoretic approach in that they can reliably extract the shape of convex service curves [28].

A drawback of the system theory in [28] is that it assumes that the measured system is time-invariant, meaning that any time-shifted packet probe experiences the same backlog or delay as the original probe. Clearly, this assumption is not satisfied in networks with random traffic load or link capacity. In this paper, we show that the system-theoretic approach to bandwidth estimation can be extended to systems with random service. We develop a stochastic bandwidth estimation scheme based on packet train probing, that is formulated in a statistical extension of the max-plus algebra of the network calculus. The offered service is expressed by $\varepsilon$-effective service curves [8], which are service curves that are allowed to violate a service guarantee with probability $\varepsilon$.

By presenting the system-theoretic formulation of available bandwidth estimation for a stochastic system, we can account for variability in a network, due to statistical properties of network traffic or transmission channels. The underlying model of the bandwidth estimation scheme can express discrete-sized probe packets, and is therefore not limited to fluid-flow traffic assumptions. An implication of our study is that the assumption of fixed-capacity FIFO links with cross traffic can be replaced by a more general network model without specific requirements on multiplexing methods. This may open the field of bandwidth estimation to network environments where FIFO assumptions are not justified.

The probing method presented in this paper employs a series of packet trains, where each packet train has a fixed inter packet gap [20, 21, 32]. This type of probing is referred to as rate scanning [28]. From these probes we estimate stationary delay distributions based on a statistical stationarity test. While, in principle, packet trains should have an infinite length to observe stationary delays, we show that, in practice, it is possible to detect stationarity with finite packet trains using statistical methods, and dynamically adapt packet trains to the required length. Using measurement results from a controlled testbed, we quantify the effect of variability on the estimated service and observe the impact of the burstiness of cross traffic on the time-scales of service availability.

The remainder of this paper is structured as follows. In Sec. 2, we discuss related work on bandwidth estimation in time-varying networks and on a generalization of bandwidth estimation in time-invariant networks. In Sec. 3, we derive a stochastic max-plus approach for estimating networks with random service. In Sec. 4, we develop our probing methodology, where we consider practical aspects such as the required length of a packet train and number of repeated measurements. In Sec. 5, we provide an experimental validation of our method. Sec. 6 gives brief conclusions.

## 2 State-of-the-art

The term *available bandwidth* denotes the capacity that is left unused by other traffic in the network, referred to as cross traffic. For a link $j$ it can be expressed for any time interval $[t, t+\tau)$ as [29]

$$\alpha_j(t, t+\tau) = \frac{1}{\tau} \int_t^{t+\tau} C_j \left(1 - u_j(x)\right) dx \,,$$

where $C_j$ is the (possibly time-varying) capacity of the link and $u_j(t) \in [0, 1]$ is its utilization by cross traffic at time $t$. For cross traffic that has a long-term average rate $\lambda_j$, the limit $\lim_{\tau \to \infty} \alpha_j(t, t+\tau) = C_j - \lambda_j$ is referred to as *average available bandwidth*. The end-to-end available bandwidth of a network path is frequently defined as the minimum of the available bandwidths of all traversed links [29, 30].

## 2.1 Bandwidth Estimation of FIFO Systems

Many bandwidth estimation tools assume a fluid time-invariant network model with FIFO scheduling. This implies that the relation between the incoming rate $r_I$ and the outgoing rate $r_O$ of a constant-rate probe at a link, referred to as rate response curve, is given by:

$$\frac{r_I}{r_O} = \begin{cases} 1 & , r_I \leq C - \lambda \\ \frac{r_I + \lambda}{C} & , r_I > C - \lambda. \end{cases}$$

Similarly, for packet pair probes, the corresponding function describing the dispersion of the probes, is called gap response curve [20, 29]. In practice, random cross traffic distorts the traffic dispersion given by response curves. To account for random cross traffic, estimation tools apply averaging [20, 37], linear regression [32], and Kalman filtering [14].

Recently, the deterministic CBR traffic model used for the response curve at a FIFO system has been extended to stochastic ones [12, 18, 29, 30, 33, 34]. A queueing theoretic framework for bandwidth estimation is analyzed in [29], where it is shown that the assumption of fluid CBR traffic generates an upper bound for the available bandwidth, and that the deviation can be resolved in principle using packet trains of infinite length. The work is extended to multi-hop networks in [30]. In [18, 33], distributions for the output gap of a probe packet pair in single-hop and multi-hop systems are derived for M|D|1 and M|G|1 queues, respectively. In [34], a bandwidth estimation tool is developed based on the distribution of the output gap. In [12], a distribution for the output gap is derived for a general arrival process in conjunction with parameter estimation for known cross traffic distributions. Fundamental limitations of active probing are analyzed in [31] based on a queueing model of a FIFO system.

Tools which do not explicitly assume FIFO scheduling, but are compatible with this assumption are, e.g., Pathchirp [36] and Pathload [21, 22]. Both tools increase their probing rate until an increase of one-way delays, or, equivalently, queueing delays of probe packets is detected. Pathload specifies the available bandwidth as a range, to capture its time-varying nature. The underlying deterministic network model is relaxed to filter out short-term fluctuations in the detection of long-term trends. As noted in [30] increasing trends can also occur due to transient effects.

## 2.2 Bandwidth Estimation of Min-Plus Linear Systems

Min-plus linear systems offer an alternative model for bandwidth estimation methods. A network is represented as a general system with traffic arrivals as input, and traffic departures as output of the system. The behavior of a time-invariant min-plus linear system is characterized by a service curve $S(t)$, which is a function that relates a system's cumulative departures $D(t)$ in an interval $[0, t)$ to its arrivals $A(t)$ by

$$D(t) = \inf_{\tau \in [0,t]} \{A(\tau) + S(t - \tau)\} =: A \otimes S(t), \tag{1}$$

where the operator $\otimes$ is defined as the convolution under the min-plus algebra. This interpretation permits a formulation of bandwidth estimation as the inversion problem of $D = A \otimes S$ for $S$ [28], where $A$ and $D$ are arrival and departure functions of probing traffic. A major advantage of a system theoretic interpretation is a straightforward extension to multi-hop settings. Given a sequence of $H$ systems where $S^h$ ($h = 1, \ldots, H$) denotes the service curve of the $h$th system, a service curve $S^{net}$ for the entire sequence of $H$ systems is given by the min-plus convolution $S^{net} = S^1 \otimes \ldots \otimes S^H$.

Using packet train arrivals $A(t) = rt$ (for a sequence of rates $r$) and measurements of $D(t)$, the maximum system backlog $B_{\max}(r) = \sup_\tau \{A(\tau) - D(\tau)\}$ can be used to compute a service curve estimate $\tilde{S}(t)$ as

$$\tilde{S}(t) = \sup_{r \geq 0} \{rt - B_{\max}(r)\} =: \mathcal{L}(B_{\max})(t),$$

where $\mathcal{L}$ denotes the Legendre transform. The above relationship can be used to justify existing packet train methods, e.g., Pathload [21]. Probing schemes using properties of the Legendre transform for min-plus linear systems have been used in [1, 19, 28].

While Eq. (1) is suitable to express the service offered by constant rate links, traffic regulators, or fair schedulers, it implies linearity under the min-plus algebra, a condition that is not satisfied by actual networks (in particular, FIFO schedulers [17]). In [28], it is argued that networks can be viewed as generally linear systems that transition to a non-linear regime when the network becomes saturated, and it is shown that the transition can be observed using suitable non-linearity tests.

Related models have been used in the context of admission control. The works [9, 40] consider strict service curves that satisfy $D(t) \geq A(\tau) + S(t - \tau)$ for any $\tau, t$ falling into the same busy period to estimate the available service. This type of service curve is also used as a basis for router parameter estimation from external [2, 39] or internal [23] measurements. While conceptually simpler, strict service curves do not extend easily to multi-hop networks.

The main limitation of the system-theoretic approach to bandwidth estimation is that the measured system must satisfy time-invariance. Dispensing with this assumption requires the development of a system-theoretic framework where traffic arrivals and departures, as well as the network service are described by random processes.

## 3 Inference of a Random Service

In this section, we develop the foundation for a stochastic bandwidth estimation methodology for networks with random service. We phrase the model in max-plus algebra [4, 10] that, unlike the min-plus approach [28], can directly operate on packet timestamps as collected by probing tools.

### 3.1 Systems with Random Service in Max-plus Algebra

Let $T_A(n)$ and $T_D(n)$, respectively, be timestamps of packet arrivals and departures, where $n = 0, 1, 2 \ldots$ is a packet index (Note that index $n$ denotes the $n+1$th packet). We use the shorthand notation $T_A(\nu, n) = T_A(n) - T_A(\nu)$. We formulate the relation of a system's arrival and departure timestamps in max-plus algebra [4, 10, 42]. The link between min-plus and max-plus systems is established by expressing $T_A(n)$ as the pseudo-inverse of $A(t)$ defined as

$$T_A(n) = \inf\{t \geq 0 : A(t) \geq n + 1\}$$

assuming unit sized packets. Considering variable sized packets requires additional notation to specify packet lengths.

Let $(A_i(t), D_i(t))$ be pairs of arrivals and corresponding departures of a system denoted by the notation $A_i(t) \mapsto D_i(t)$. To establish a relation to min-plus linear systems observe that $\min\{A_1(t), A_2(t)\} \mapsto \min\{D_1(t), D_2(t)\}$ becomes $\max\{T_{A_1}(n), T_{A_2}(n)\} \mapsto \max\{T_{D_1}(n), T_{D_2}(n)\}$ by pseudo-inversion, whereas the mapping $A(t) + \nu \mapsto D(t) + \nu$ translates to shift-invariance $T_A(n+\nu) \mapsto T_D(n+\nu)$ in max-plus algebra. Conversely, the second condition required to achieve max-plus linearity $T_A(n) + \tau \mapsto T_D(n) + \tau$ implies time-invariance $A(t+\tau) \mapsto D(t+\tau)$ in min-plus algebra.

In analogy to Eq. (1), a system is max-plus linear and shift-invariant [4] (implying min-plus linearity and time-invariance) if and only if

$$T_D(n) = \max_{\nu \in [0,n]} \{T_A(\nu) + T_S(n - \nu)\} =: T_A \otimes T_S(n),$$

where $\otimes$ denotes the max-plus convolution and $T_S(n-\nu)$ is the system's service curve in max-plus algebra that specifies the amount of time spent on serving $n - \nu + 1$ packets [10]. For non-linear systems, service curves cannot provide an exact mapping of arrivals to departures, however, they can provide bounds of the form $T_D(n) \leq T_A \otimes T_S(n)$.

Dispensing with the assumption of shift-invariance, we substitute $T_S(n - \nu)$ by the bivariate function $T_S(\nu, n)$, which can express a random service experienced by a sequence of packets $(\nu, n)$. A definition of bivariate service curve is

$$T_D(n) = \max_{\nu \in [0,n]} \{T_A(\nu) + T_S(\nu, n)\} =: T_A \otimes T_S(n). \tag{2}$$

Note that the shift-varying definition retains max-plus linearity.

In the stochastic network calculus, random service can be modeled by $\varepsilon$-effective service curves that express a non-random shift-invariant bound on the available service that may be violated with a defined probability $\varepsilon$ [8]. In the max-plus algebra, an $\varepsilon$-effective service curve $T_S^\varepsilon(n)$ can be defined to specify a service guarantee of the form

$$\mathsf{P}\left[T_D(n) \leq \max_{\nu \in [0,n]} \{T_A(\nu) + T_S^\varepsilon(n-\nu)\}\right] > 1 - \varepsilon, \tag{3}$$

where $\varepsilon$ is a small violation probability. The following lemma links the definitions in Eqs. (2) and (3). It views varying service as a random process and specifies $\varepsilon$-effective service curves as a stationary bound.

**Lemma 1.** *Given a system with bivariate service curve $T_S(\nu, n)$ as in Eq. (2). Any function $T_S^\varepsilon(n)$ that satisfies*

$$\mathsf{P}\left[T_S(\nu, n) \leq T_S^\varepsilon(n-\nu), \forall \nu\right] > 1 - \varepsilon$$

*for $n \geq 0$ is an $\varepsilon$-effective service curve in the sense of Eq. (3) of the system.*

*Proof.* Consider a sample path $T_S^\omega(\nu, n)$ of $T_S(\nu, n)$ and fix $n \geq 0$. If $T_S^\omega(\nu, n) \leq T_S^\varepsilon(n-\nu)$ for all $\nu \in [0, n]$, it follows from the monotonicity of max-plus convolution that

$$T_D(n) = T_A \otimes T_S^\omega(n) \leq T_A \otimes T_S^\varepsilon(n).$$

Since, by assumption, the condition $T_S^\omega(\nu, n) \leq T_S^\varepsilon(n-\nu)$ for all $\nu \in [0, n]$ holds at least with probability $1 - \varepsilon$, the claim follows. □

### 3.2 Estimation of Effective Max-plus Service Curves

In this section, we show how an $\varepsilon$-effective service curve can be obtained from constant rate packet train probes $T_A(n) = n/r$. To this end, we define the delay of packet $n$ as $W(n) = T_D(n) - T_A(n)$, and phrase the delay as a function of the probing rate, that is, $W(r, n) = T_D(n) - n/r$. The steady-state delay for $n \to \infty$ is abbreviated by $W(r)$. We define the $(1 - \varepsilon)$-percentile of the delay distribution by

$$W^\varepsilon(r, n) = \inf\{x \geq 0 : \mathsf{P}[W(r,n) \leq x] > 1 - \varepsilon_W\},$$

where we use $\varepsilon_W$ to denote the $(1-\varepsilon_W)$-percentile of the delay distribution. With these definitions, Th. 1 provides the foundation for a packet train based estimation method.

**Theorem 1.** *Given a system with bivariate service curve $T_S(\nu, n)$ as in Eq. (2). For all $n \geq 0$ the function*

$$T_S^\varepsilon(n) = \inf_{r \geq 0}\left\{\frac{n}{r} + W^\varepsilon(r)\right\} =: \mathcal{F}(W^\varepsilon)(n)$$

*is an $\varepsilon$-effective service curve in the sense of Eq. (3) of the system with violation probability $\varepsilon_S = \sum_r \varepsilon_W$.*

*Proof.* From $W(n) = T_D(n) - T_A(n)$ and Eq. (2) it follows that

$$W(n) = \max_{\nu \in [0,n]}\{T_S(\nu, n) - T_A(\nu, n)\}. \tag{4}$$

The maximum in Eq. (4) implies that $W(n) \geq T_S(\nu, n) - T_A(\nu, n)$ for all $\nu \in [0, n]$, permitting us to write

$$T_S(\nu, n) \leq T_A(\nu, n) + W(n), \forall \nu.$$

Inserting $T_A(n) = n/r$ and using the delay percentile yields

$$\mathsf{P}\left[T_S(\nu, n) \leq \frac{n-\nu}{r} + W^\varepsilon(r, n), \forall \nu\right] > 1 - \varepsilon_W.$$

By application of the union bound it follows that

$$\mathsf{P}\left[T_S(\nu,n) \le \inf_{r \ge 0}\left\{\frac{n-\nu}{r} + W^\varepsilon(r,n)\right\}, \forall \nu\right] > 1 - \sum_r \varepsilon_W.$$

With Lem. 1 we obtain that $T_S^\varepsilon(n-\nu)$ defined as

$$T_S^\varepsilon(n-\nu) = \inf_{r \ge 0}\left\{\frac{n-\nu}{r} + W^\varepsilon(r,n)\right\} \quad (5)$$

for all $\nu \in [0,n]$ is an $\varepsilon$-effective service curve with violation probability $\varepsilon_S = \sum_r \varepsilon_W$. Letting $n \to \infty$ and inserting the steady-state delay $W^\varepsilon(r)$ completes the proof. $\square$

Similar to the relation of backlog and min-plus service curves established by the Legendre transform, see e.g., [16] for details, Th. 1 relates delay and max-plus service curves using the transform $\mathcal{F}$, which shares characteristics of a Legendre transform. To see this, we substitute $r = 1/s$ and $W^\varepsilon(r) = -V^\varepsilon(s)$, permitting us to rewrite the result of Th. 1 as

$$T_S^\varepsilon(n) = \inf_{s \ge 0}\{sn - V^\varepsilon(s)\},$$

which is the (concave) Legendre transform of $V$. From the properties of the Legendre transform it is known that $T_S^\varepsilon(n)$ is a concave function, and, moreover, if $V^\varepsilon(s)$ is concave, it holds that

$$V^\varepsilon(s) = \inf_{n \ge 0}\{sn - T_S^\varepsilon(n)\}.$$

In other words, for concave functions, the transform is its own inverse. This establishes $W^\varepsilon(r)$ as a dual characterization of systems, equivalent to a characterization by $T_S^\varepsilon(n)$.

Th. 1 gives rise to a method for service curve estimation for linear networks with random service using packet train probes sent at different rates $r$. Each packet train provides a sample of the steady-state delay $W(r)$ from which an estimate of the delay percentiles, denoted by $\widetilde{W}^\varepsilon(r)$, is obtained. Applying the transform from Th. 1, we can compute an estimate of an $\varepsilon$-effective service curve as $\widetilde{T}_S(n) = \mathcal{F}(\widetilde{W}^\varepsilon)(n)$.

*Example:* We consider the estimation method for a system consisting of a random On-Off server. In each time slot, the server performs an independent Bernoulli trial and forwards a packet with probability $p = 0.1$. We use constant rate probes consisting of 100 packets sent at different rates. Fig. 1 shows the $\varepsilon$-effective service curves with $\varepsilon = 10^{-3}$ computed with Th. 1. As indicated by the thin dash-dotted lines in Fig. 1, each probing rate $r$ contributes a linear segment with slope $1/r$ and axis intercept $W^\varepsilon(r)$ to the service curve estimate. Hence, the resolution of the piecewise linear estimate increases by adding probing rates. At the same time, adding probing rates increases the violation probability due to the use of the union bound in Th. 1. An analytical lower and upper bound of the best possible service curve are included in the graph as a reference. The lower bound is computed as the number of time slots required such that $n$ packets are forwarded with probability at least $1 - \varepsilon$. The number of time slots required to forward one packet follows a geometric distribution and the sum of $n$ independent geometric random variables has negative binomial distribution. The lower bound follows as the $(1-\varepsilon)$-percentile of the negative binomial distribution. The upper bound is computed from the lower bound by application of the union bound. We denote by $\lim_{n \to \infty} n/T_S^\varepsilon(n)$ the *limiting rate* of the service curve. For comparison, we show the service curve of a server that has a constant rate equal to the average available bandwidth. Note that the limiting rate of the service curve is equal to the average available bandwidth.

Since our method uses steady-state delays to obtain a service curve estimate, the convergence to a steady-state is vital. To this end, the following lemma proves the existence of the steady-state delay distribution as long as the probing rate does not exceed the limiting rate of the service curve. We use the notation $X =_d Y$ to mean that two random variables $X$ and $Y$ are equal in distribution, i.e., $\mathsf{P}[X \le a] = \mathsf{P}[Y \le a]$.

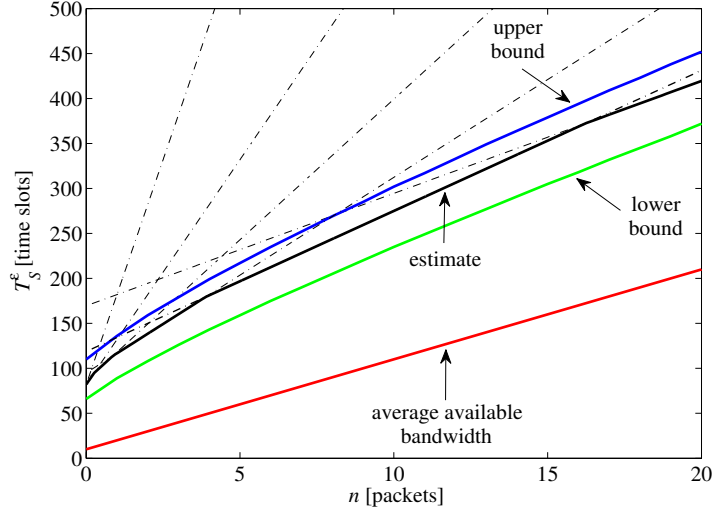

Figure 1: Service curves of an On-Off server. The service curve estimate is composed of linear segments, that are each obtained by a single probing rate. Analytical lower and upper bounds are included for comparison. The limiting rate of the service curve converges to the average available bandwidth. Compared to the average available bandwidth, the service curve provides significant details on the time-scales of service availability.

**Lemma 2.** *Given arrivals $T_A(n)$ at a system with bivariate service curve $T_S(\nu, n)$. If $T_A(0,n) =_d T_A(\nu, n+\nu)$ and $T_S(0,n) =_d T_S(\nu, n+\nu)$ for all $n, \nu$, the delay is stochastically increasing. Extend the processes $T_A(\nu, n)$ and $T_S(\nu, n)$ from $0 \le \nu \le n < \infty$ to $-\infty < \nu \le n < \infty$. If for all $n$ it holds that*
$$\limsup_{\nu \to \infty} \frac{T_A(n-\nu, n)}{\nu} > \limsup_{\nu \to \infty} \frac{T_S(n-\nu, n)}{\nu}$$
*almost surely, the delay converges in distribution to a random variable $W$.*

The lemma extends Lem. 9.1.4 in [10] from backlogs at a constant rate server phrased in min-plus algebra to delays at a server with a varying service in max-plus algebra. The proof closely follows [10].

*Proof.* From Eq. (4) it follows for any $x$ that
$$\mathsf{P}[W(n+1) \ge x]$$
$$= \mathsf{P}\left[\max_{\nu \in [0, n+1]} \{T_S(\nu, n+1) - T_A(\nu, n+1)\} \ge x\right]$$
$$\ge \mathsf{P}\left[\max_{\nu \in [0, n]} \{T_S(\nu+1, n+1) - T_A(\nu+1, n+1)\} \ge x\right].$$

With $T_A(\nu+1, n+1) =_d T_A(\nu, n)$ and $T_S(\nu+1, n+1) =_d T_S(\nu, n)$ for all $n, \nu$ the last line equals $\mathsf{P}[W(n) \ge x]$. Hence, $\mathsf{P}[W(n+1) \ge x] \ge \mathsf{P}[W(n) \ge x]$ which proves that the delay $W(n)$ is stochastically increasing.

From the second assumption of Lem. 2 it follows that for any $n$ there exists a finite random variable
$$N = \max\{\nu \ge 0 : T_A(n-\nu, n) \le T_S(n-\nu, n)\}.$$

Thus, $T_S(n-\nu, n) < T_A(n-\nu, n)$ holds for all $\nu > N$. Moreover, since $T_S(n-\nu, n)$ increases with $\nu \ge 0$ we have $T_S(n-\nu, n) \le T_S(n-N, n)$ for all $0 \le \nu \le N$. Combining the two statements and using that $T_A(n-\nu, n)$ for $\nu \geq 0$ and $T_S(n-N, n)$ are non-negative yields

$$T_S(n-\nu, n) - T_A(n-\nu, n) \leq T_S(n-N, n)$$

for all $\nu \geq 0$ and hence

$$\max_{\nu \geq 0}\{T_S(n-\nu, n) - T_A(n-\nu, n)\} \leq T_S(n-N, n).$$

With $\max_{\nu \geq 0}\{T_S(n-\nu, n) - T_A(n-\nu, n)\} = W(n)$ from Eq. (4) it follows for any $x$ that

$$\sup_n\{\mathsf{P}[W(n) \geq x]\} \leq \sup_n\{\mathsf{P}[T_S(n-N, n) \geq x]\}.$$

Since $N$ is finite and $W(n)$ is stochastically increasing there exists a finite random variable $W$ such that

$$\lim_{n \to \infty} \mathsf{P}[W(n) \geq x] = \sup_n\{\mathsf{P}[W(n) \geq x]\} = \mathsf{P}[W \geq x]$$

completes the proof of the second statement. □

## 3.3 Connection to Min-plus Stochastic Network Calculus

In the remainder of this section, we show how the estimation method that is established by Th. 1 can be mirrored in the min-plus algebra, where the backlog takes the place of the delay. We connect the two approaches and state how a service curve estimate in max-plus algebra can be transformed into a service curve in min-plus algebra. The majority of methods from the network calculus have been formulated using the min-plus algebra.

Similar to Eqs. (2) and (3) a definition of bivariate service curve in min-plus algebra is [10]

$$D(t) = \inf_{\tau \in [0,t]}\{A(\tau) + S(\tau, t)\} =: A \otimes S(t), \quad (6)$$

which characterizes min-plus linear time-varying systems, and a definition of $\varepsilon$-effective service curve is [8]

$$\mathsf{P}\left[D(t) \geq \inf_{\tau \in [0,t]}\{A(\tau) + S^\varepsilon(t-\tau)\}\right] > 1 - \varepsilon, \quad (7)$$

respectively. The following lemma links the two definitions.

**Lemma 3.** *Given a system with bivariate service curve as in Eq. (6). Any function $S^\varepsilon(t)$ that satisfies*

$$\mathsf{P}\big[S(\tau, t) \geq S^\varepsilon(t-\tau), \forall \tau\big] > 1 - \varepsilon$$

*for $t \geq 0$ is an $\varepsilon$-effective service curve in the sense of Eq. (7) of the system.*

*Proof.* Consider a sample path $S^\omega(\tau, t)$ of $S(\tau, t)$ and fix $t \geq 0$. If $S^\omega(\tau, t) \geq S^\varepsilon(t-\tau)$ for all $\tau \in [0, t]$, it follows that

$$D(t) = A \otimes S^\omega(t) \geq A \otimes S^\varepsilon(t).$$

Since $S^\omega(\tau, t) \geq S^\varepsilon(t-\tau)$ for all $\tau \in [0, t]$ holds at least with probability $1 - \varepsilon$, the lemma is proven. □

Let $B^\varepsilon(r)$ be the $(1-\varepsilon)$-percentile of the steady-state backlog of a system with constant bit rate arrivals with rate $r$. We use $\varepsilon_B$ to denote the $(1-\varepsilon_B)$-percentile of the backlog distribution. Th. 2 shows how a service curve can be estimated from $B^\varepsilon(r)$.

**Theorem 2.** *Given a system with bivariate service curve as in Eq. (6). For all $t \geq 0$ the function*

$$S^\varepsilon(t) = \sup_{r \geq 0}\{rt - B^\varepsilon(r)\} = \mathcal{L}(B^\varepsilon)(t)$$

*is an $\varepsilon$-effective service curve in the sense of Eq. (7) of the system with violation probability $\varepsilon_S = \sum_r \varepsilon_B$.*

*Proof.* From $B(t) = A(t) - D(t)$ and Eq. (6) we have $B(t) = \sup_{\tau \in [0,t]}\{A(\tau,t) - S(\tau,t)\}$. The supremum implies that $B(t) \geq A(\tau,t) - S(\tau,t)$ for all $\tau \in [0,t]$ permitting us to solve for $S(\tau,t) \geq A(\tau,t) - B(t)$ for all $\tau \in [0,t]$. Inserting $A(\tau,t) = r(t-\tau)$ and using the backlog percentile yields

$$\mathsf{P}\left[S(\tau,t) \geq r(t-\tau) - B^\varepsilon(r,t), \forall \tau\right] > 1 - \varepsilon_B.$$

By application of the union bound it follows that

$$\mathsf{P}\left[S(\tau,t) \geq \sup_{r \geq 0}\{r(t-\tau) - B^\varepsilon(r,t)\}, \forall \tau\right] > 1 - \sum_r \varepsilon_B.$$

Letting $S^\varepsilon(t-\tau) = \sup_{r \geq 0}\{r(t-\tau) - B^\varepsilon(r,t)\}$ we conclude with Lem. 3 that $S^\varepsilon(t-\tau)$ is an $\varepsilon$-effective service curve with violation probability $\varepsilon_S = \sum_r \varepsilon_B$. Finally, we let $t \to \infty$ and insert the steady-state backlog $B^\varepsilon(r)$. □

The min-plus network calculus models arrivals as infinitely divisible. To establish a connection to the max-plus packet model we use the concept of a packetizer $P^L(x)$ [10, 26]. The packetizer delays bits to convert a fluid data flow into a packetized one with cumulative packet lengths $L(n)$. Assuming unit sized packets the definition of packetizer simplifies to $P^L(x) = \lfloor x \rfloor$ [26]. If the arrivals $A(t)$ are packetized, it holds that $\lfloor A(t) \rfloor = A(t)$. Given packetized arrivals the following lemma shows that $\lfloor S^\varepsilon(t) \rfloor$ provides a service curve guarantee for the packetized departures $\lfloor D(t) \rfloor$.

**Lemma 4.** *Given a system with $\varepsilon$-effective service curve as in Eq. (7) and packetized arrivals $A(t) = \lfloor A(t) \rfloor$. It holds that*

$$\mathsf{P}\bigl[\lfloor D(t) \rfloor \geq \lfloor A(t) \rfloor \otimes \lfloor S^\varepsilon(t) \rfloor\bigr] > 1 - \varepsilon.$$

*Proof.* From Eq. (7) we obtain that

$$\mathsf{P}\bigl[D(t) \geq A \otimes S^\varepsilon(t)\bigr] \leq \mathsf{P}\bigl[\lfloor D(t) \rfloor \geq \lfloor A \otimes S^\varepsilon(t) \rfloor\bigr]$$

since generally if $D(t) \geq A \otimes S^\varepsilon(t)$, it holds that $\lfloor D(t) \rfloor \geq \lfloor A \otimes S^\varepsilon(t) \rfloor$. Since the arrivals are assumed to be packetized we can substitute $A(t) = \lfloor A(t) \rfloor$ such that

$$\lfloor A \otimes S^\varepsilon \rfloor = \lfloor \lfloor A \rfloor \otimes S^\varepsilon \rfloor = \lfloor A \rfloor \otimes \lfloor S^\varepsilon \rfloor.$$

By insertion of the second line into the first line it follows that $\lfloor S^\varepsilon(t) \rfloor$ is an $\varepsilon$-effective service curve for the packetized departures $\lfloor D(t) \rfloor$. □

The proof of Th. 3 uses Lem. 5 that relates the backlog of a system to the delay.

**Lemma 5.** *Given a system with arrivals $T_A(n)$ and departures $T_D(n)$. Assume the system serves arrivals in order. The backlog equals*

$$B(T_D(n)) = A(T_D(n) - W(n), T_D(n)),$$

*where $A(\tau,t)$ are the cumulative arrivals in $[\tau,t)$ and $W(n)$ is the delay of packet $n$.*

*Proof.* From the definition of backlog we have

$$B(T_D(n)) = A(T_D(n)) - D(T_D(n)).$$

Since the arrivals are served in order it holds that $D(T_D(n)) = A(T_A(n))$ and it follows by substitution that

$$B(T_D(n)) = A(T_A(n), T_D(n)).$$

Using the definition of delay $W(n) = T_D(n) - T_A(n)$ yields

$$B(T_D(n)) = A(T_D(n) - W(n), T_D(n)),$$

which completes the proof. □

We use Lem. 5 to relate percentiles of the backlog and delay to each other, e.g., $B^\varepsilon = rW^\varepsilon$ for constant rate arrivals with rate $r$, where we denote by $B$ and $W$ the steady-state backlog and delay for $n \to \infty$. We note that Little's law, i.e., $\mathsf{E}[B] = \lambda \mathsf{E}[W]$ for arrivals with average rate $\lambda$, can be recovered from Lem. 5. To see this, assume stationary arrivals, i.e., $A(t) = A(\tau, t+\tau)$ for all $\tau \geq 0$. Letting $n \to \infty$ and taking expectations we obtain $\mathsf{E}[B] = \mathsf{E}[A(W)]$, where we used the stationarity of $A$. Conditioning on $W = w$ it follows that

$$\mathsf{E}[B] = \mathsf{E}\big[\mathsf{E}[A(w)|W=w]\big].$$

We decompose $A(w) = A(t) + A(t, 2t) + \ldots A(w-t, w)$ into $w/t$ increments and let $t \to 0$. Under the earlier assumption of stationarity, all increments are identically distributed. Further on, we assume that the increments are statistically independent of each other and of the service. It follows that the increments are independent of $W$, i.e., future packet arrivals do not depend on the delay observed by the current packet. Since the expected value of a sum is the sum of the expected values of each summand we obtain

$$\mathsf{E}[B] = \mathsf{E}\big[W/t\, \mathsf{E}[A(t)|W=w]\big] = \mathsf{E}[W]\, \mathsf{E}[A(t)]/t.$$

Since the increments of $A(t)$ are identically distributed $\mathsf{E}[A(t)]/t$ is constant for all $t > 0$. Letting $\lambda = \mathsf{E}[A(t)]/t$ we arrive at Little's law $\mathsf{E}[B] = \lambda \mathsf{E}[W]$.

The following theorem links the max-plus service curve estimate to the packetized min-plus service curve estimate by pseudo-inversion.

**Theorem 3.** *Consider $T_S^\varepsilon(n)$ and $S^\varepsilon(t)$ from Th. 1 and Th. 2, respectively, and assume the system forwards arrivals in order. It holds that*

$$\lfloor S^\varepsilon(t) \rfloor = \min\{n \geq 0 : T_S^\varepsilon(n+1) > t\} =: (T_S^\varepsilon)^{-1}(t).$$

*Proof.* By insertion of $T_S^\varepsilon(n)$ from Th. 1 we have

$$(T_S^\varepsilon)^{-1}(t) = \min\left\{n \geq 0 : \inf_{r \geq 0}\left\{\frac{n+1}{r} + W^\varepsilon(r)\right\} > t\right\}.$$

After some reordering we arrive at

$$(T_S^\varepsilon)^{-1}(t) = \min\left\{n \geq 0 : n > \sup_{r \geq 0}\{rt - rW^\varepsilon(r)\} - 1\right\}.$$

Instantiating Lem. 5 with $A(\tau, t) = r(t - \tau)$ yields the backlog $B(T_D(n)) = rW(n)$. Letting $n \to \infty$ and taking percentiles we obtain $B^\varepsilon(r) = rW^\varepsilon(r)$ and it follows that

$$(T_S^\varepsilon)^{-1}(t) = \min\left\{n \geq 0 : n > \sup_{r \geq 0}\{rt - B^\varepsilon(r)\} - 1\right\}.$$

Substituting $\sup_r \{rt - B^\varepsilon(r)\}$ by $S^\varepsilon(t)$ using Th. 2

$$(T_S^\varepsilon)^{-1}(t) = \min\{n \geq 0 : n > S^\varepsilon(t) - 1\} = \lfloor S^\varepsilon(t) \rfloor$$

completes the proof. □

## 4 Probing Methodology

Sec. 3 establishes a method that computes a service curve of a link, a router, or an entire network path from steady-state delay percentiles $W^\varepsilon(r)$ obtained from packet train probes. We specify constant rate packet train probes by $\langle N, R, I \rangle$, where $N$ is the train length, $R$ is the set of rates that are probed, and $I$ is the number of repeated measurements. In principle, the steady-state delay $W(r)$ is obtained from the last packet of an infinitely long packet train ($N \to \infty$). Also, the tail distribution of the delay requires an infinite number of repeated measurements ($I \to \infty$). In this section, we use statistical methods to obtain delay estimates using small packet trains and a small number of repeated measurements. We present methods for selecting the parameters of packet train probes, which adapt to the characteristics of cross traffic in the network.

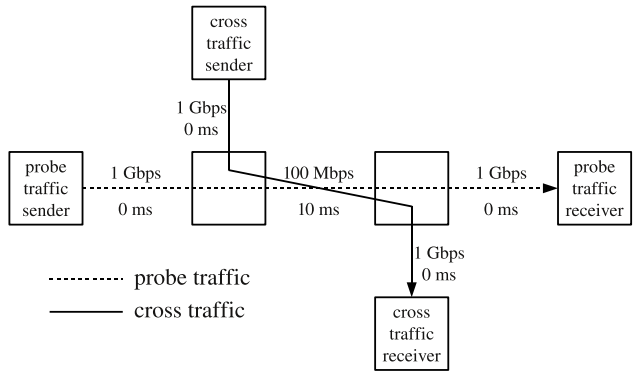

Figure 2: Dumbbell topology with 100 Mbps bottleneck link.

For an experimental evaluation of our methods, we use the Emulab testbed [41], which offers controlled experiments using real networking equipment. We consider a dumbbell topology as shown in Fig. 2, where cross traffic and probe traffic share a 100 Mbps bottleneck link. The capacities and propagation delays of the links are specified in the figure. We consider different types of linear and non-linear schedulers, including priority, fair queueing, and FIFO, and different buffer sizes at the bottleneck link. The default setting uses priority scheduling with high priority to cross traffic, and a large buffer size (of $10^6$ packets). Cross traffic has a mean rate of 50 Mbps, and consists of equally spaced packet bursts whose size follows a truncated Exponential or Pareto distribution. The average size of a packet burst of both distributions is 1500 Byte and the shape parameter of the Pareto distribution is 1.5. We use D-ITG [3] for traffic generation. D-ITG truncates datagrams at 64 kByte to conform to the maximum IP payload. IP fragmentation is used to create packets with a size of at most 1500 Byte. Rude/Crude [24] is employed to emit packet train probes consisting of packets of 1500 Byte size. The NTP derivate Chrony [11] is used for time synchronization.

## 4.1 Selection of Probing Rates

The selection of the set of probing rates presents a tradeoff. On the one hand, adding probing rates improves the estimate of the $\varepsilon$-effective service curves, since each rate contributes an additional linear segment (see Fig. 1). On the other hand, since Th. 1 computes the violation probability by an application of the union bound, adding probing rates increases the violation probability $\varepsilon$. We use an algorithm that seeks to find a small set of characteristic rates that contribute significantly to the service curve. The algorithm is a combination of a binary increase and a binary search algorithm, similar to the rate selection procedure in [21].

The algorithm has as parameter $r_{acc}$ to specify the desired rate resolution. Binary increase starts at $r_1 = r_{acc}$. As long as the probes at rate $r_i$ measure a finite delay percentile the rate is doubled. The first rate $r_i$ at which the test fails is used to start a binary search in the interval $[r_{i-1}, r_i]$ using the same test criterion. Each additional rate that is probed halves the interval. Once the interval achieves the target accuracy $r_{acc}$ the rate scan is terminated. Let $\tilde{r}$ be the largest rate that achieves a finite delay percentile. As an example if $\tilde{r} = 50$ Mbps and $r_{acc} = 4$ Mbps the algorithm probes at rates $r_i = 4, 8, 16, 32, 64, 48, 56,$ and 52 Mbps and stops at the interval $[48, 52]$.

The number of rates probed by the binary increase/binary search algorithm is $2\lfloor \log_2(\tilde{r}/r_{acc}) \rfloor + 2$. The binary increase algorithm requires $\lfloor \log_2(\tilde{r}/r_{acc}) \rfloor + 2$ steps until it first exceeds $\tilde{r}$ and the binary search performs another $\lfloor \log_2(\tilde{r}/r_{acc}) \rfloor$ steps to ensure the target accuracy.

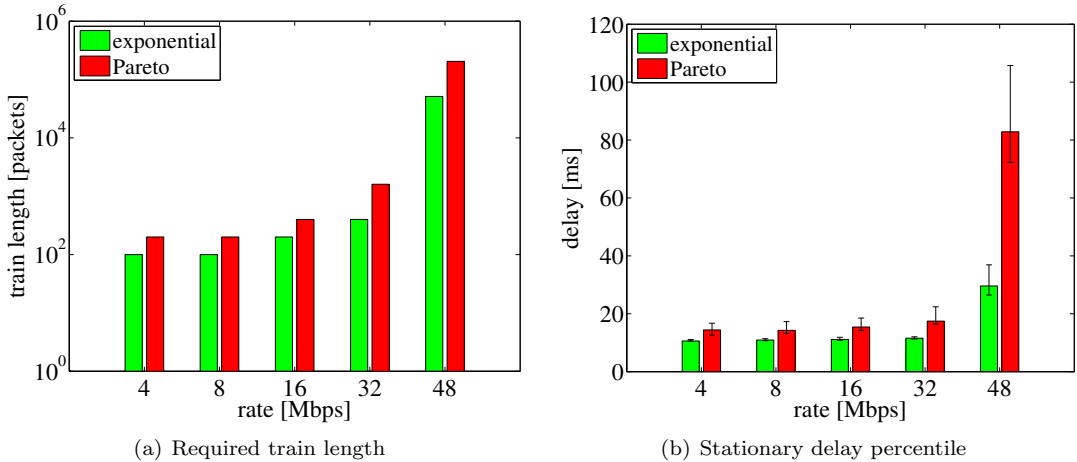

(a) Required train length

(b) Stationary delay percentile

Figure 3: Train lengths that permit observing stationary delays at a 100 Mbps link with 50 Mbps cross traffic and respective 0.95 delay percentiles with 0.95 confidence intervals. Bursty cross traffic requires longer trains to detect stationarity. The required train length increases drastically when the probing rate approaches the average available bandwidth of 50 Mbps.

## 4.2 Estimation of Steady-State Delays

We now discuss how to estimate steady-state delays using finite length packet train probes. We employ a statistical test that detects stationarity of a time series, and use it to define a procedure that adapts the train length to the variability of cross traffic. Finally, based on the stationarity test we formulate a heuristic to estimate a service curve and its limiting rate while reducing the amount of probes.

### 4.2.1 Stationarity Test

Th. 1 uses steady-state delays to compute a service curve. While Lem. 2 states that the steady-state delay distribution exists as long as the rate of probe arrivals does not exceed the limiting rate of the service curve, reaching the steady-state requires infinitely long packet trains. To determine the steady-state delay from a finite length packet train, we use a statistical test that detects if the delays $W(n)$ of a sequence of probe packets $n \in [0, N-1]$ satisfy stationarity. If stationarity is detected, we use the delay of the last packet of a packet train as an estimate of the steady-state delay. The delay values of all other packets from the same packet train are discarded due to possible correlations. If stationarity cannot be detected, we set the delay estimate to infinity. We repeat the measurement $I$ times for each probing rate $r$ to measure the $(1-\varepsilon)$-percentile of the delay $W^\varepsilon(r)$. Note that due to the infimum in Th. 1, delay percentiles of infinity do not contribute to the service curve estimate.

To detect stationarity of the delay series observed by a packet train, we use the unit root test from Elliot, Rothenberg and Stock (ERS) [15, 35]. The *ERS test* is based on an auto-regressive moving average model. The null hypothesis of the test is that the data series has a unit root, which implies non-stationarity. The ERS test returns a single value referred to as *ERS statistic*. If the ERS statistic falls below a critical value, the null hypothesis is rejected and stationarity is assumed.

### 4.2.2 Adaptive Train Length

Since the minimal train length that permits detecting stationarity is not known a priori we define a procedure that adaptively increases the train length. When the ERS test indicates non-stationarity for a share of more than $\varepsilon$ of the packet trains sent at a certain rate, then either the stationary $(1-\varepsilon)$-delay percentiles cannot be achieved at this probing rate, or the train length is too short to reliably detect stationarity. To asses whether increasing the train length can help we consult the trend of the ERS statistic. We compute the ERS statistic for the first half of the train and for the entire train. If the ERS statistic decreases, we interpret this as an indication that longer trains may achieve stationarity. We refer to this test as the *trend test* that is passed by a packet train if its ERS statistic decreases. If the majority of the packet trains sent at a certain rate passes the trend test, we double the train length and carry out the measurements at this rate anew. The measurements repeat this procedure until either stationarity is achieved or the trend test fails.

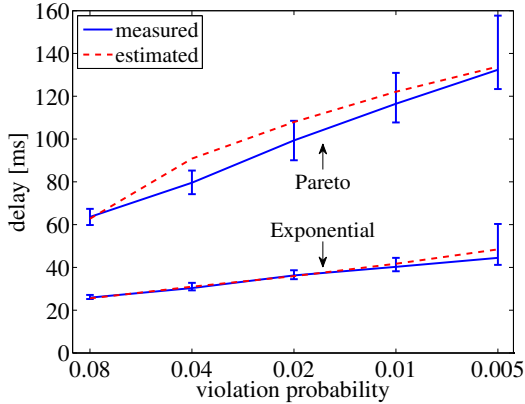

Figure 4: Direct estimation of $W^\varepsilon(r)$ from 2000 delay samples compared to an estimate obtained from the POT method using 250 samples only. The POT method shows a good fit.

First, we analyze the train length that is required to achieve stationarity. We come back to a full evaluation of the trend test in Sec. 4.2.4. Fig. 3(a) shows the train lengths that permit detecting stationarity for a share of at least $1-\varepsilon_W = 0.95$ of $I = 250$ packet trains sent at different probing rates and for different types of cross traffic each. The probing rates are chosen according to the algorithm from Sec. 4.1 and the train lengths are adapted as described above starting at a minimum train length of 100 packets. The results show that the required train length is very sensitive to the distribution of cross traffic and to the probing rate. The required train length increases sharply, when the probing rate approaches the average available bandwidth of 50 Mbps.

### 4.2.3 Tail Distribution

The computation of $\varepsilon$-effective service curves from Th. 1 uses the $(1-\varepsilon)$-percentiles of the delay. To obtain an estimate of the delay percentiles $\widetilde{W}^\varepsilon(r)$, we repeat packet train measurements $I$ times for each probing rate, resulting in $I$ delay samples per rate. The delays observed by different packet trains are assumed to be independent if probes are emitted at random start times (see [5, 38] for a discussion). We quantify the accuracy of the estimated delay percentiles using confidence intervals (which, for percentiles, are computed from the binomial distribution, see e.g., [25]). Fig. 3(b) displays the 0.95 delay percentiles and corresponding 0.95 confidence intervals as observed by the packet trains from Fig. 3(a). Our measurements show that the width of the confidence intervals increases when the probing rate approaches the limiting rate and when cross traffic is more bursty. Fig. 3(b) provides the input to estimate a service curve from Th. 1.

For small violation probabilities $\varepsilon$ the direct extraction of delay percentiles becomes inefficient requiring a large number of iterations to obtain reliable estimates. In this case we apply the peaks over threshold (POT) method from extreme value theory, see e.g., [6], to predict tail probabilities of the delays. Given a number of samples the POT method parameterizes a generalized Pareto distribution such that it fits the tail distribution of the samples above a certain threshold. Using the POT method we obtain $\widetilde{W}^\varepsilon(r)$ and related confidence intervals for small $\varepsilon$ from the generalized

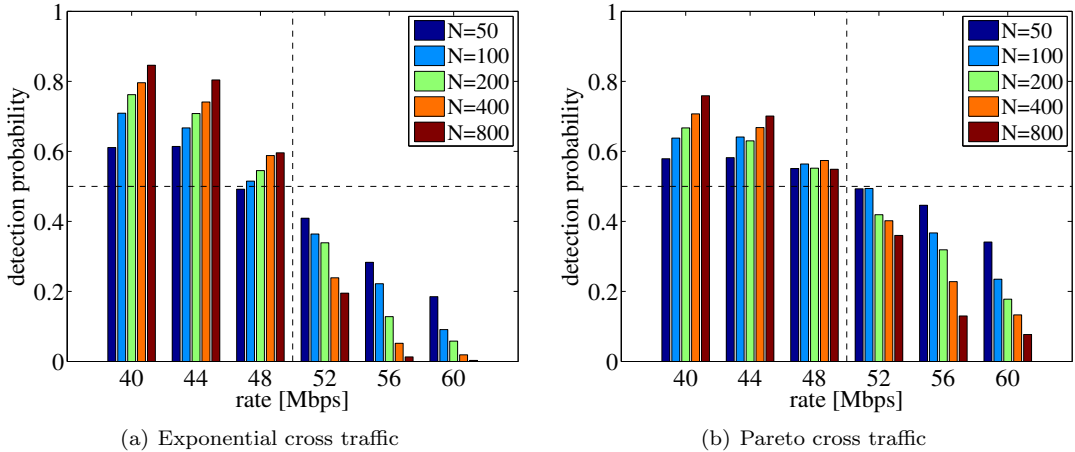

(a) Exponential cross traffic  (b) Pareto cross traffic

Figure 5: Short packet trains are used to determine the trend of the ERS statistic. If a decreasing trend is detected, the existence of a stationary delay distribution is assumed. The probability that a train of length $N$ passes the trend test, i.e., detects a decreasing trend, is shown. The limiting rate of the network is 50 Mbps. The majority of the trains below (above) 50 Mbps are classified correctly, i.e., pass (fail) the test.

Pareto distribution that is parameterized using a limited set of observations only. The POT method assumes independent and identically distributed samples. Moreover, the distribution of the samples must be in the domain of attraction of an extreme value distribution [6]. We apply the tests from [13, 27] to our measurement data to verify this condition before using the POT method.

Fig. 4 shows the application of the POT method to derive $(1-\varepsilon)$-percentiles of the delay for small $\varepsilon$. We compare delay percentiles that are directly extracted from 2000 independent packet trains sent at a rate of 48 Mbps to percentiles estimated from the POT method. We configured the POT method to use only 250 delay samples and a threshold of 0.9. The results show a good fit for Exponential as well as for Pareto cross traffic.

### 4.2.4 Limiting Rate Estimation

The limiting rate of a service curve can be determined as the largest probing rate that observes steady-state delays, see Lem. 2. To this end, packet trains of sufficient length are required as analyzed in Fig. 3(a). In this subsection, we develop and evaluate a heuristic that detects the limiting rate from a minimum of probing traffic. We use short packet trains of a fixed length $N$. If a packet train passes either the ERS test or otherwise the trend test from Sec. 4.2.2, we assume that a steady-state delay exists for the current probing rate. If both tests fail, we assume the delay is infinite. Using this outcome we immediately continue the binary increase/binary search algorithm with the next rate. However, different from the procedure in Sec. 4.2.2, we do not increase the train length to actually observe stationary delays.

We start with an evaluation of the train length and its impact on the fidelity of the trend test. For the network in Fig. 2 the limiting rate is 50 Mbps, i.e., trains with a rate below (above) 50 Mbps should pass (fail) the trend test. Fig. 5 depicts the relative frequency of trains that pass the trend test. In the figure, each bar aggregates the results from 1000 repeated experiments for train lengths of $50-800$ packets sent at rates $40-60$ Mbps. Note that the majority of the trains are classified correctly, providing an empirical justification for the test. The likelihood of a correct classification increases when the train length is increased. The accuracy decreases when the probing rate approaches the limiting rate. The accuracy is generally lower for burstier cross traffic.

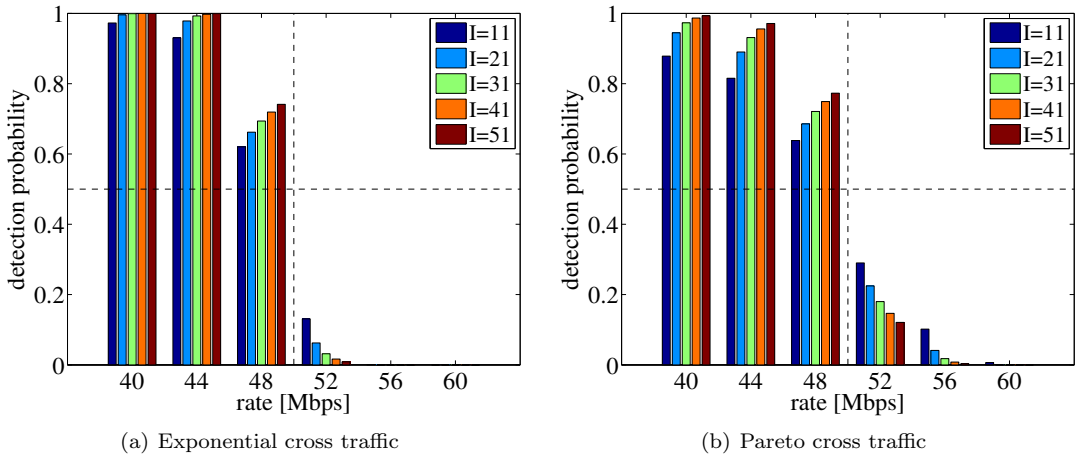

(a) Exponential cross traffic      (b) Pareto cross traffic

Figure 6: The test from Fig. 5 is improved by a majority decision using $I$ repeated measurements. Results are shown for a train length of $N = 200$ packets. With increasing $I$ the probability to pass (fail) the test reaches one (zero). If the limiting rate is approached, significantly larger $I$ are required.

To improve the robustness of the test we repeat the packet trains for a given probing rate $I$ times and conduct a majority decision. Assuming independence we compute the probability that the majority of the trains passes the test from Fig. 5 via the binomial distribution. Fig. 6 shows the results for a train length of $N = 200$ packets and $I = 11 - 51$ repeated measurements. $I$ is chosen to be an odd number to facilitate an unambiguous majority decision. The probability of correct classification increases in $I$ and approaches one for large $I$. The closer the limiting rate has to be tracked, the more iterations are required.

#### 4.2.5 Service Curve Estimation

While the trend test can use short packet train probes to estimate the limiting rate, short packet trains do not generally permit observing stationary delays to compute a service curve estimate from Th. 1. In that case we can still obtain a service curve estimate from Eq. (5) that has, however, a restricted domain. Using the procedure from Sec. 4.2.4 we measure the delay of the last packet indexed $N - 1$ of each train and compute a service curve estimate $T_{\tilde{S}}^{\varepsilon}(n)$ from Eq. (5) for all $n \in [0, N-1]$.

We compare service curves obtained with packet trains of restricted length $N$ to those obtained by a priori unrestricted, adaptive train lengths that observe stationary delays. We choose a target resolution of $r_{acc} = 4$ Mbps and obtain $1 - \varepsilon_W = 0.95$ percentiles of the delay from $I = 250$ repeated measurements. As Fig. 7 shows all service curve estimates closely track the limiting rate of 50 Mbps, which is indicated in the figure by dashed reference lines. Recall that the estimates are composed of linear segments with a slope that is proportional to the reciprocal value of the probing rate and an axis intercept that equals the delay observed at this rate. We note that the estimates coincide initially regardless of the packet train length. The respective segments are obtained from probing rates that are well below the limiting rate, where even short trains observe stationary delays. Once the probing rate approaches the limiting rate, delay estimates obtained from short trains underestimate the stationary delay distribution, resulting in a more optimistic service curve estimate.

### 4.3 Non-linear and Lossy Networks

Our estimation method is derived under the assumption that network elements such as links, queues, and schedulers can be modeled by lossless linear systems. While numerous networks, such as links with a varying capacity or priority schedulers satisfy linearity, FIFO schedulers are a notable exception. However, as shown in [17], a FIFO scheduler operates in a linear regime if the aggregated data rate of probe and cross traffic is less than the capacity of the system; it turns into a non-linear regime when the FIFO scheduler is overloaded. This observation permits the application of linear systems to FIFO schedulers as long as the capacity is not fully utilized [28]. Even though our stationarity test ensures that the probing traffic does not exhaust the capacity of the network in the long run, bursty cross traffic may cause short term violations. The assumption of lossless systems can be relaxed under the max-plus algebra, where we model dropped packets as incurring an infinite delay, i.e., if packet $n$ has been dropped, we set $T_D(n) = \infty$. As a consequence, probing rates experiencing a packet loss ratio of $\varepsilon$ or more do not contribute to the service curve estimate.

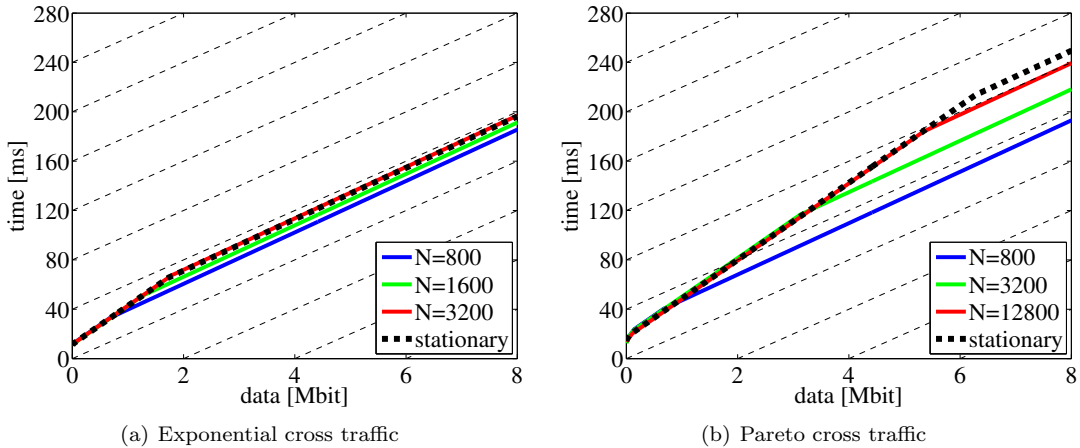

(a) Exponential cross traffic  (b) Pareto cross traffic

Figure 7: Service curve estimates from trains of length $N$ compared to unrestricted trains that observe stationary delays. The estimates coincide at the beginning where small probing rates contribute. Moreover, all estimates reveal the limiting rate of 50 Mbps indicated by the dashed reference lines. Shorter trains measure, however, smaller delays resulting in more optimistic estimates.

Fig. 8 shows service curve estimates obtained for priority, fair, and FIFO scheduling at the bottleneck link. We also include results for a FIFO scheduler with a small buffer size of 200 packets, which will result in moderate packet losses. The probing parameters are $r_{acc} = 4$ Mbps, $I = 250$, $\varepsilon_W = 0.05$, and $N = 800$ as in Fig. 7. For all scenarios, the axis intercept slightly above 10 ms matches the propagation delay of the bottleneck link (see Fig. 2). The service curve estimate for the fair scheduler is a straight line since the scheduler allocates a fair share of 50 Mbps to the probe traffic regardless of the burstiness of cross traffic. The service curve estimate for low priority probes at a priority scheduler, on the other hand, is sensitive to the type of cross traffic. Note that the estimate for the FIFO scheduler with and without packet losses are very close, indicating that our estimation method deals well with packet losses. In the case of Pareto cross traffic and with small buffers packet losses result in an optimistic service curve estimate since large cross traffic bursts are cut off by the small buffer.

## 5 Comparative Evaluation

We present a comparison with related methods and tools for bandwidth estimation. Measurement results are obtained for the network topology in Fig. 2, with FIFO scheduling, and parameters as discussed in Sec. 4.

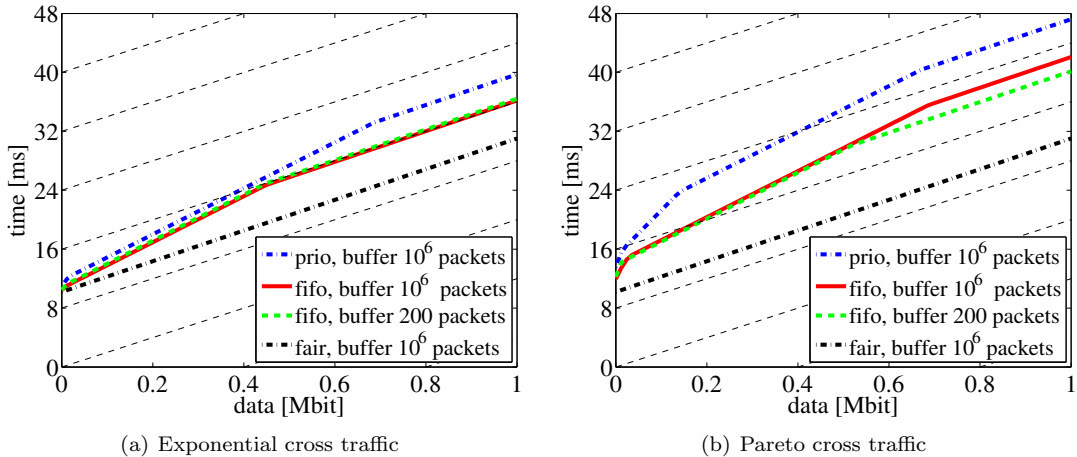

(a) Exponential cross traffic  (b) Pareto cross traffic

Figure 8: Service curve estimates for priority, fair, and FIFO scheduling with large and small buffer. The estimates for FIFO lie between priority scheduling, where cross traffic bursts are served first, and fair scheduling, which achieves a fair rate of 50 Mbps regardless of the variability of cross traffic.

## 5.1 Service Curve Estimation

We first provide a comparison with the rate scanning method from [28], which results in an estimate for a deterministic service curve using min-plus linear time-invariant system theory. The method in [28] employs constant rate packet train probes with a fixed length (800 packets), and increments the rate between successive trains by a constant amount (8 Mbps) up to a maximum rate of 120 Mbps. Since the system theory in [28] is derived using a min-plus algebra, we convert our results for comparison. Also, we limit the adaptively varied train length of our method to at most 800 packets, and use a target accuracy of $r_{acc} = 4$ Mbps. We aggregate results that are obtained from 200 iterations. Recall that our method uses repeated measurements to obtain $(1 - \varepsilon)$-percentiles, here $\varepsilon_W = 0.05$, and corresponding 0.95 confidence intervals of the delay to derive a single estimate of an $\varepsilon$-service curve. The rate scanning method [28] generates a separate estimate in each iteration, and uses the estimates from all iterations to compute the mean and 0.95 confidence intervals.

Fig. 9 shows the service curve estimates with confidence intervals obtained with the two methods for Exponential and Pareto cross traffic. The buffer limit is set to $10^6$ packets since the deterministic approach from [28] does not account for packet losses. The estimates for Exponential cross traffic, Fig. 9(a), are comparable for both methods, however, a comparison with the diagonal reference lines shows that the deterministic service curve overestimates the limiting rate of 50 Mbps. For Pareto cross traffic, shown in Fig. 9(b), the mean of the estimates for the deterministic service curve deviates noticeably from the limiting rate. The figure makes evident that the $\varepsilon$-service curve provides a more reliable estimate with tight confidence intervals.

## 5.2 Limiting Rate Estimation

We next compare our method to selected bandwidth estimation tools of which some are frequently used as benchmark methods. We include the following methods in our comparison: Pathload, IGI/PTR, Spruce, and dietTOPP. For Pathload, which denotes a range of values for the available bandwidth, the lower and upper bounds are labeled *PL lb* and *PL ub*, respectively. Our method is labeled as *SCest*. For all tools we use the default parameters. Our method is configured to achieve a target accuracy of 1 Mbps. We use packet trains of 100 packets and perform 11 iterations for each rate. For comparison, Pathload uses trains of length 100 and performs 12 iterations. For each evaluated method, we perform 100 repeated measurements. The results are displayed in

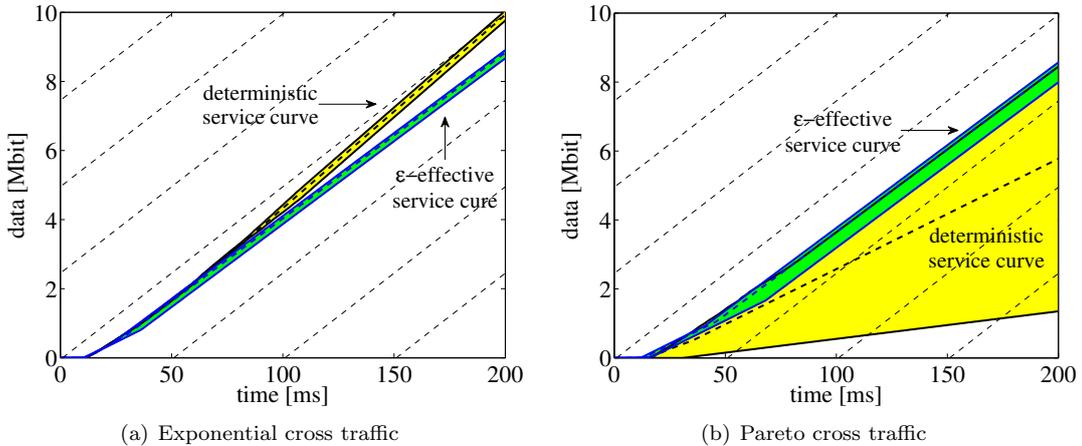

(a) Exponential cross traffic
(b) Pareto cross traffic

Figure 9: Service curve estimates (dashed lines) and respective 0.95 confidence intervals (shaded areas) from our method compared to the approach from [28]. In case of mildly variable Exponential cross traffic both methods produce comparable results. If cross traffic is more bursty, i.e., Pareto, the deterministic approach [28] produces, however, highly variable estimates. In contrast, our method succeeds in providing an estimate with tight confidence intervals that closely approaches the limiting rate.

Fig. 10, where for each method we show the median, the 0.25 and 0.75 percentiles, and the 0.05 and 0.95 percentiles as box-plots. Moreover, Fig. 10 depicts the average available bandwidth of 50 Mbps as a horizontal line. The buffer limit is 200 packets causing moderate packet losses. We also conducted experiments with larger buffers yielding similar results.

Figs. 10(a) and 10(b), respectively, show the bandwidth estimates for Exponential and Pareto cross traffic. Our SCest method compares favorably with other methods, providing a lower bound of the limiting rate, while some methods tend to overestimate the available bandwidth. The strength of our method becomes evident in the case of highly bursty Pareto cross traffic.

## 6  Conclusion

In this paper, we developed a new foundation for bandwidth estimation of networks with random service. We used the framework of the stochastic network calculus to derive a method that estimates an $\varepsilon$-effective service curve from steady-state delay percentiles observed by packet train probes. The service curve model extends to networks of nodes and is not restricted to specific schedulers, such as FIFO. It characterizes service availability at different time scales and recovers the well-known average available bandwidth as its limiting rate. We used methods from statistics, specifically a stationarity test, to determine the parameters of packet train probes. We found that cross traffic variability and the target accuracy of the estimate have a significant impact on the amount of probes required. We presented and evaluated heuristic methods that can effectively reduce the volume of probe traffic. We showed a comparison with related methods and tools.

## Acknowledgements

The work of R. Lübben and M. Fidler is supported by an Individual grant and in part by an Emmy Noether grant from the German Research Foundation (DFG), respectively. The work of J. Liebeherr is supported in part by an NSERC Strategic Project.

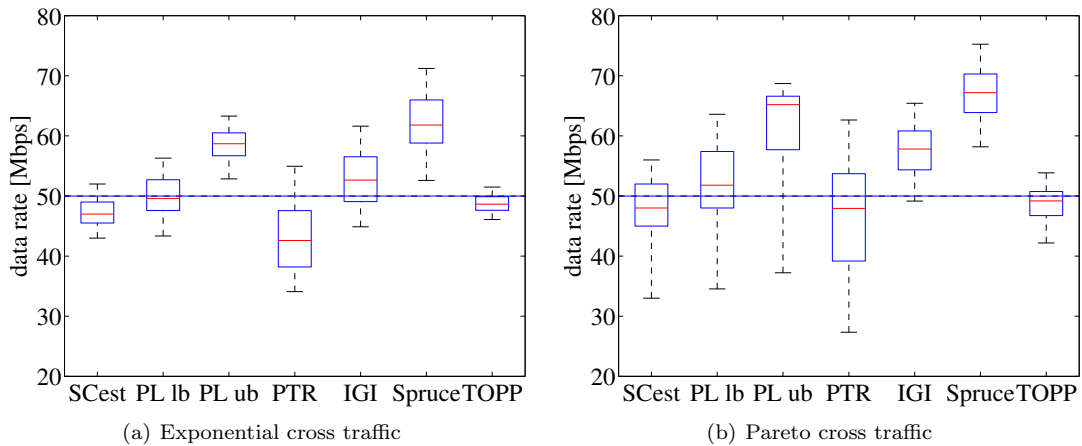

Figure 10: Comparison of bandwidth estimation tools. For each tool the median, 0.25 and 0.75 percentiles, and 0.05 and 0.95 percentiles from 100 trials are given. For both, Exponential and Pareto cross traffic our method provides a lower bound that matches the available bandwidth well.